\documentclass[journal]{IEEEtran}
\newtheorem{theorem}{Theorem}
\newtheorem{proposition}[theorem]{Proposition}

\newtheorem{cor}[theorem]{Corollary}

\newcommand{\Z}{{\mathbb Z}}
\newcommand{\F}{{\mathbb F}}

\usepackage{amsmath,amssymb,amsfonts}

\begin{document}
\title{Decomposition of bent generalized Boolean functions}

\author{Lin Sok, MinJia Shi and Patrick Sol\'{e}%
\thanks{Lin Sok, School of Mathematical Sciences, Anhui University, Hefei, Anhui, 230601 and Department of Mathematics, Royal University of Phnom Penh, Cambodia. {\tt sok.lin@rupp.edu.kh}}

\thanks{MinJia Shi, Key Laboratory of Intelligent Computing \& Signal Processing,
Ministry of Education, Anhui University
No. 3 Feixi Road, Hefei
Anhui Province 230039, P. R. China, National Mobile Communications Research Laboratory, Southeast University, 210096, Nanjing,  P. R. China
and School of Mathematical Sciences of Anhui University, Hefei, 230601, P. R. China. {\tt smjwcl.good@163.com}}

\thanks {Patrick Sol\'{e}, CNRS/LAGA, University of Paris 8, 93 526 Saint-Denis, France, {\tt patrick.sole@telecom-paristech.fr} }

}

\maketitle

\begin{abstract}
A one to one correspondence between regular generalized bent functions from $\F_2^n$ to $\Z_{2^m},$ and $m-$tuples of Boolean bent functions is established.
This correspondence maps self-dual
(resp. anti-self-dual) generalized bent functions to $m-$tuples of self-dual (resp. anti self-dual) Boolean bent functions.
An application to the classification of regular generalized bent functions under the extended affine group is given.
\end{abstract}

\medskip

{\bf Keywords: Boolean functions, generalized bent functions, Walsh Hadamard transform}


\section{Introduction}

Bent functions have been a popular topic in difference sets, symmetric Cryptography, and Coding theory since their introduction by Rothaus in 1976\cite{C1}. They are a building brick of streamcipher systems, and offer optimal resistance to fast correlation attacks and affine approximation attacks \cite{C1}. Two recent books are dedicated
to this important concept \cite{M,T}. In recent years, a theory of {\em generalized} Boolean function, in the sense that the domain range is no longer $\F_2,$ but an arbitrary
$\Z_q,$ has arisen \cite{Stan,ST,T}.

In this paper we extend the theory of decomposition of quaternary Boolean functions of \cite{SSS}, from $q=4$ to $q$ an arbitrary power of $2.$ First, we establish the decomposition
for  regular bent functions, then apply it to self-dual bent functions, which are regular by definition. As a byproduct, we give a necessary existence
condition for regular generalized bent functions. More importantly, we use this decomposition to classify regular bent functions ($q$ a power of $2$) under the action of the extended
affine group when the number of variables is at most $4.$

The material is organized as follows. The next section collects the notations and definitions that we need in the rest of the paper. Section 3 develops the decomposition technique
for generalized bent functions. Section 4 classifies regular quaternary bent function under the action of the affine group. Section 5 concludes the article.

\section{Definitions and notation}
\subsection{Bent functions}
A {\em Boolean function} in $n$ variables is any function from $\F_2^n$ to $\F_2.$ The set of all $2^{2^n}$ such functions is denoted by ${\cal B}_n.$
The {\em sign function} of $f$ is defined as $F(x)=(-1)^{f(x)}.$
The {\em Walsh-Hadamard}(WHT) transform $W_f(u)$ of the Boolean function $f,$ evaluated in a point $u$ of the domain $\F_2^n,$ is defined as $W_f(u)=\sum_{x\in \F_2^n}(-1)^{x.u}F(x).$
Alternatively, in matrix terms, if $F$ is viewed as a column vector the matrix of the WHT is the Hadamard matrix $H_n$ of Sylvester type,
which we now define by tensor products. Let
\[
H:=\left(\begin{array}{cc}
1 & 1 \\
1& -1
\end{array}\right).
\]

Let $H_n:=H^{\otimes n}$ be the $n$-fold tensor product of $H$ with itself and
${\cal H}_n:=H^{\otimes n}/2^{n/2},$ its normalized version. Recall the Hadamard property
\[
H_nH_n^T=2^nI_{2^n},
\]
where we denote by $I_M$ the $M$ by $M$ identity matrix. With these notations $W_f(u)=H_nF.$
A function $f \in {\cal B}_n,$ is said to be {\em bent} if $W_f(u)=\pm 2^{n/2}$ for all $u \in \F_2^n.$ By integrality reasons such functions only exist for even $n.$
If $f$ is bent its {\em dual} function is defined as that element $\widehat{f}$ of ${\cal B}_n$ such that its sign function,
 henceforth denoted by $\widehat{f},$ satisfies $\widehat{f}=\frac{W_f(u)}{2^n}$. If, furthermore, $f=\widehat{f}$, then $f$ is
\emph{self-dual bent} \cite{CDPS}. Similarly, if $f=\widehat{f}+1$ then $f$ is \emph{anti-self-dual bent}\cite{CDPS}.
Thus if $f$ is self-dual bent its sign function is an eigenvector of $H_n$ associated to the eigenvalue $2^{n/2}.$ Likewise, if $f$ is anti self-dual bent its sign function is an eigenvector of $H_n$ associated to the eigenvalue $-2^{n/2}.$
\subsection{Generalized bent functions}
A {\em generalized Boolean function}(gBF) in $n$ variables is any function from $\F_2^n$ to $\Z_q,$ for some integer $q.$ In this work we shall focus on the case $q=2^m,$ for some integer $m>1.$
The set of all such gBFs will be denoted by ${\cal GB}_n.$ The (complex) {\em sign function} of $f$ is defined as $F(x)=(\omega)^{f(x)},$
where $\omega$ stands for a complex root of unity of order $2^m.$
The  {\em Walsh-Hadamard} transform $H_f(u)$ of the Boolean function $f,$ evaluated in a point $u$ of the domain $\F_2^n,$ is defined as $H_f(u)=\sum_{x\in \F_2^n}(-1)^{x.u}F(x).$ In matrix terms $H_f(u)=H_nF.$
A function $f \in {\cal GB}_n,$ is said to be {\em bent} if $|H_f(u)|=2^{n/2}$ for all $u \in \F_2^n.$
A bent gBF is said to be {\em regular} if there is an element $\widehat{f}$ of ${\cal Q}_n,$ such that its sign function satisfies $H_f(u)=2^{n/2}\widehat{f}.$
If, furthermore, $f=\widehat{f}$, then $f$ is
\emph{self-dual bent}. Similarly, if $f=\widehat{f}+2^{m-1},$ then $f$ is \emph{anti-self-dual bent}.
\section{Decomposition}
By standard facts on cyclotomic polynomials, we know that the degree of $\omega$ over the rationals is $k=2^{m-1}$.

{\bf Definition:} A system of $2^s$ boolean functions $f_0,\cdots,f_{2^s-1},$ with respective sign functions $F_0,\cdots,F_{2^s-1},$ is said to have the {\em Hadamard property} if
$$H_s (F_0,\cdots,F_{2^s-1})^{\top}$$ is equal to $\pm$ some column of $H_s.$ For instance, the condition is automatically verified for $s=1.$ It becomes non trivial as soon as the
possible number of columns of length $2^s,$ that is $2^{2^s}$ exceeds twice the number of columns of $H_s$ that is $2^{s+1}.$ The latter condition is equivalent to $s>1.$

\begin{theorem}\label{theo:dec1} If the sign function of the regular bent gBF $f$ is $\omega^f =\sum_{i=0}^{k-1}a_i \omega^ i,$ then the $k$ BF $G_i$ for $i=0,\cdots,k-1$ defined by
$$(G_0,\cdots,G_{k-1})^{\top}=H_{m-1}(a_0,\cdots,a_{k-1})^{\top}  $$ are bent BF with the Hadamard property, and so is the system of their duals. Conversely, given $k$ BF $G_0,\cdots,G_{k-1},$ with the Hadamard property, with duals also with Hadamard property, the gBF of sign function $\sum_{i=0}^{k-1}a_i \omega^ i$
with the $a_i$'s are defined by the above system is regular bent.
\end{theorem}

\begin{pf}
Note first that the $a_i$'s taking values $0,\pm 1$ and with supports partitionning $\F_2^n,$  the values taken by the $G_i$'s are in $\pm 1.$
Thus, the $G_i$'s can be regarded as sign functions of BF $g_i$'s say. Since for a given element in the domain exactly one $a_i$ is nonzero with value $\pm1$ we see that
the system of the $G_i$'s affords the Hadamard property.
Because $f$ is a regular gBF we can write
$H_n\omega^f =2^{n/2}\omega^g,$ say, with $\omega^g=\sum_{i=0}^{k-1}b_i \omega^ i,$
with the $b_i$'s taking values $0,\pm 1$ and with supports partitionning $\F_2^n.$
Since $\{1, \omega, \cdots, \omega^{k-1}\}$ is an integral basis of $\Z[\omega]$ we can write the $k$ equalities $H_n a_i=2^{n/2}b_i.$ Taking linear combinations we get
$H_nG_i=2^{n/2}e_iH_{m-1}(b_0,\cdots,b_{k-1})^{\top},$ where $e_i$ denotes the element $i$ of the canonical basis in $k$ dimensions, viewed as row vector. By the same argument as above for the $a_i$'s but with the $b_i$'s we see that $e_iH_{m-1}(b_0,\cdots,b_{k-1})^{\top},$ takes values in $\pm 1.$ Hence the $k$ BF $g_i$'s are bent and the system of their duals affords the Hadamard property.
 Reversing the order of the above considerations yields the converse. The Hadamard property of the system of the $G_i$'s shows, using $H_{m-1}H_{m-1}^{\top}=kI_k,$ that the supports of the
 $a_i$'s partition $\F_2^n,$ and thus that the functions $F$ defined by $F=\sum_{i=0}^{k-1}a_i \omega^ i,$ is indeed a complex sign function of the form $i^f,$ for some gBF $f.$
 This function is seen to be regular bent by taking linear combinations of the equalities $H_nG_i=2^{n/2}(-1)^{\widehat{g_i}},$ and using the fact that the system of the duals
 also satisfy the Hadamard property.
\end{pf}

\begin{cor} There is no regular bent $\Z_{2^m}$-valued gBF in odd number of variables.
\end{cor}
\begin{pf}
The function $a_0$ is a classical bent function in $n$ variables like $f.$ It is well-known since Rothaus that there is no bent function in odd number of variables \cite{C1}.
\end{pf}

We now specialize this decomposition to the case of self-dual gBFs and self-dual BFs. Note that self-dual gBFs are regular.
\begin{theorem} If the sign function of the self-dual bent gBF $f$ is $\omega^f =\sum_{i=0}^{k-1}a_i \omega^ i,$ then the $k$ self-dual BFs $G_i$ for $i=0,\cdots,k-1$ defined by
$$(G_0,\cdots,G_{k-1})^{\top}=H_{m-1}(a_0,\cdots,a_{k-1})^{\top}  $$ are bent BF with the Hadamard property. Conversely, given $k$ BF $G_0,\cdots,G_{k-1},$ with the Hadamard property,the gBF of sign function $\sum_{i=0}^{k-1}a_i \omega^ i$
where the $a_i$'s are defined by the above system is self-dual bent.
\end{theorem}
\begin{pf}
Note first that the $a_i$'s taking values $0,\pm 1$ and with supports partitionning $\F_2^n,$  the values taken by the $G_i$'s are in $\pm 1.$
Thus, the $G_i$'s can be regarded as sign functions of BFs $g_i$'s say. Since for a given element in the domain exactly one $a_i$ is nonzero with value $\pm1$ we see that
the system of the $G_i$'s affords the Hadamard property.
Because $f$ is a self-dual gBF we can write
$H_n\omega^f =2^{n/2}\omega^f.$
Since $\{1, \omega, \cdots, \omega^{k-1}\}$ is an integral basis of $\Z[\omega]$ we can write the $k$ equalities $H_n a_i=2^{n/2}a_i.$ Taking linear combinations we get
$H_nG_i=2^{n/2}e_iH_{m-1}(a_0,\cdots,a_{k-1})^{\top},$ where $e_i$ denotes the element $i$ of the canonical basis in $k$ dimensions, viewed as row vector. By the same argument as above  we see that the quantity $e_iH_{m-1}(b_0,\cdots,b_{k-1})^{\top},$ takes values in $\pm 1.$ Hence the $k$ BF $g_i$'s are self-dual bent.
 Reversing the order of the above considerations yields the converse. The Hadamard property of the system of the $G_i$'s shows, using $H_{m-1}H_{m-1}^{\top}=kI_k,$ that the supports of the
 $a_i$'s partition $\F_2^n,$ and thus that the functions $F$ defined by $F=\sum_{i=0}^{k-1}a_i \omega^ i,$ is indeed a complex sign function of the form $i^f,$ for some gBF $f.$
 This function is seen to be self-dual bent by taking linear combinations of the equalities $H_nG_i=2^{n/2}(-1)^{{g_i}}.$
\end{pf}
\section{Classification}
In this section, we classify all quaternary regular bent functions ($q=4$), of degree at most $4,$ under the action of the extended affine group.
The equivalence of two regular bent functions is defined as follows.
\begin{proposition}\label{prop:equiv}Let $f$ be a quaternary regular bent function in $n$ variables. Then $g(x)=f(xM+a)+c$, where $M\in GL(n,2)$, $a\in \F_2^n$ and $c\in \Z_4$ is also regular bent.
\end{proposition}
\begin{pf}
Assume that $g(x)=f(xM+a)+c$ as in the proposition. Then, for any $u\in \F_2^n,$\\
$W_g(u)=i^c(-1)^{a.u{(M^{-1})}^\top}W_f(u{(M^{-1})}^\top)=i^c(-1)^{a.u{(M^{-1})}^\top}2^{\frac{n}{2}}i^{\widehat{f}(u)}=2^{\frac{n}{2}}i^{\widehat{g}(u)},$
with $\widehat{g}(u)=\widehat{f}(u)+c+2(a.u{(M^{-1}))^\top},$ and
where the first equality is obtained from the substitution $x=xM+a$ in the WHT of $g$ and the second equality is due to
the fact that $f$ is regular bent.
\end{pf}\\
{\bf Remark:} Two functions $f$ and $g$ defined in Proposition \ref{prop:equiv} are said to be affinely equivalent.
It is well known that two binary bent functions are EA-equivalent if $g(x)=f(xM+a)+bx+c$, where $M\in GL(n,2)$, $a,b\in \F_2^n$ and $c\in \F_2$. However for our quaternary regular bent functions, there is only restricted EA-equivalence, that is, $b=0$.\\

By applying our decomposition technique of  Theorem \ref{theo:dec1}, we can now classify all quaternary regular bent functions upto four variables and we give all representatives in Table
1 below.
\begin{theorem}Up to affine equivalence, there are $2,7$ non-equivalent quaternary regular bent functions in $2,4$. The number of quaternary reqular bent functions is the  square of that of binary case and more precisely there are $8^2, 896^2,(3502\times 13888)^2$ in $2,4,6$ variables respectively.
\end{theorem}
\begin{table}\label{Table:1}\caption{Quaternary regularbent functions in two and four variables}
$$
\begin{array}{|c|c|}
\hline
 \text{Representative from equivalence class}& Size\\
\hline
2101&16\\
2000&48\\
\hline
\text{Number of quaternary regular bent functions in two variables}&64\\
\hline
 2 0 0 0 2 0 2 2 2 0 0 0 0 2 0 0 &1792\\
 3 1 0 0 3 1 2 2 3 1 1 1 1 3 1 1 &80640\\
 2 1 0 1 2 0 2 2 3 0 0 1 0 2 1 1 &129024\\
 3 0 0 1 2 0 2 2 3 1 0 0 0 3 0 1 &215040\\
 3 1 0 0 3 0 3 2 2 1 0 1 1 3 0 0 &322560\\
 2 1 0 1 2 1 2 3 2 1 0 1 0 3 0 1 &26880\\
 2 0 1 1 2 0 2 2 2 0 0 0 0 2 1 1 &26880\\
\hline
\text{Number of quaternary regular bent functions in four variables}&802816\\
\hline
\end{array}
$$
\end{table}
\section{Conclusion}
In this article we have decomposed bent generalized Boolean functions with values in $\Z_{2^m},$ as a function of certain systems of $2^{m-1}$ bent Boolean functions.
The natural question that arises would be to replace $\Z_{2^m},$ by $\Z_{p^m},$ for odd $p,$ or even $\Z_{q},$ for an arbitrary $q.$ However, in these cases, the condition of Hadamard type seems to be never satisfied in view of the existence of power of $\omega$ that more than one nonzero component on the basis of $\Z[\omega].$ For instance, if $q=6,$ we have
$\omega^2=\omega-1,$ when a basis of $\Z[\omega]$ is $\{1,\omega\}.$


\bigskip

\begin{thebibliography}{11}
\bibitem{C1}C. Carlet, {\it Boolean Functions for Cryptography and Error  Correcting Codes}, chapter in {\it Boolean methods and models} Cambridge University Press (Peter Hammer and Yves Crama eds), to appear.
\bibitem{C2}C. Carlet, {\it On the secondary constructions of resilient and bent
functions}, Proceedings of the Workshop on Coding, Cryptography and
Combinatorics 2003, K. Feng,  H. Niederreiter and C. Xing Eds., pp. 3--28, Progress in Comp. Sc. and Appl. Logic, Birkh\"auser Verlag, 2004.
\bibitem{CDPS}C. Carlet, Lars Eirik Danielsen, Matthew G. Parker, P. Sol\'e, Self dual bent functions, Int. J. Inform. and Coding Theory 1(4), pp. 384--399, 2010.
\bibitem{DPS} Lars Eirik Danielsen, Matthew G. Parker, P. Sol\'e, Self dual bent functions, Springer Lecture Notes in Computer Science, LNCS 5921, pp. 418--432, 2009.
\bibitem{J} Janusz, G.J.
`Parametrization of self-dual codes by orthogonal matrices,'
{\sl Finite Fields Appl.,} Vol.~13, No.~3,(2007) 450--491.
\bibitem{M} S. Mesnager, {\it Bent Functions}, Springer, Berlin (2016).
\bibitem{SSS}M. Shi, L. Sok, P. Sol\'e, Classification and Construction of quaternary self-dual bent functions, Proceedings of SETA 2016.
\bibitem{ST} P. Sol\'e, N. Tokareva Connections between quaternary and binary bent functions // Cryptology ePrint Archive, Report 2009/544. http://eprint.iacr.org (eng)
\bibitem{Stan} P. Stanica, T. Martinsen, S. Gangopadhyay, B.K. Singh, Bent and generalized bent functions, Des. Codes and Crypto 69 (2013) 77-94.
\bibitem{T}N. Tokareva, {\it Bent Functions: Results and Applications to Cryptography}, Academic Press, New-York (2015).


\end{thebibliography}
\end{document}